\documentclass[aps,prl,twocolumn,times,amsmath,amssymb,superscriptaddress,floatfix]{revtex4}

\usepackage{times}
\usepackage{float}
\usepackage{amsmath}
\usepackage{amsthm}
\usepackage{amssymb}
\usepackage{amsbsy}
\usepackage{graphicx}
\usepackage{pstricks}
\usepackage{color}
\usepackage{cancel}
\usepackage{subfigure} 
\usepackage{subfloat}
\usepackage{epstopdf}    
\usepackage{enumerate}
\usepackage{mathrsfs} 
\usepackage{multirow} 
\usepackage{ulem}
\usepackage{wasysym}
\usepackage{sidecap}

\begin{document}

\title{A spin liquid with pinch-line singularities on the pyrochlore lattice}

\author{Owen Benton}
%
\affiliation{Okinawa Institute of Science and Technology Graduate University, Onna-son,
Okinawa 904-0495, Japan}

\author{L. D. C. Jaubert}

\affiliation{Okinawa Institute of Science and Technology Graduate University, Onna-son,
Okinawa 904-0495, Japan}

\author{Han Yan}

\affiliation{Okinawa Institute of Science and Technology Graduate University, Onna-son,
Okinawa 904-0495, Japan}

\author{Nic Shannon}
%
\affiliation{Okinawa Institute of Science and Technology Graduate University, Onna-son,
Okinawa 904-0495, Japan}

\maketitle

\noindent {\bf Abstract}\\

The mathematics of gauge theories lies behind many of the most profound advances 
in physics in the last 200 years, from Maxwell's theory of electromagnetism to Einstein's 
theory of general relativity. 
More recently it has become clear that gauge theories also emerge in condensed matter,
a prime example being the spin ice materials which host an emergent electromagnetic
gauge field. 
In spin ice, the underlying gauge structure is revealed by the presence 
of pinch-point singularities in neutron-scattering measurements. 
Here we report the discovery of a spin liquid where 
the low-temperature physics is 
naturally described by the fluctuations of a tensor field
with a continuous gauge freedom.
This gauge structure underpins an unusual form of spin correlations, 
giving rise to pinch-line singularities
--- line-like analogues of the pinch-points observed in spin ice. 
Remarkably, these features may already have been
observed in the pyrochlore material Tb$_2$Ti$_2$O$_7$.
\\

\noindent {\bf Introduction} \\


Gauge symmetries are paramount in the understanding
of many of the most fundamental theories of physics.
Recent decades have seen an increasing 
appreciation of the role of gauge theories
in condensed matter physics, emerging 
from the long-wavelength description 
of the collective behaviour of electrons. 
Emergent gauge theories have proved particularly 
important in the study of spin liquids -- strongly fluctuating,
disordered magnetic states, the description of 
which lies beyond the familiar territory of Landau 
theory \cite{lee92, wen02, moessner03, balents10,henley10, gingras14}.

\begin{figure}
\centering
\includegraphics[width=\columnwidth]{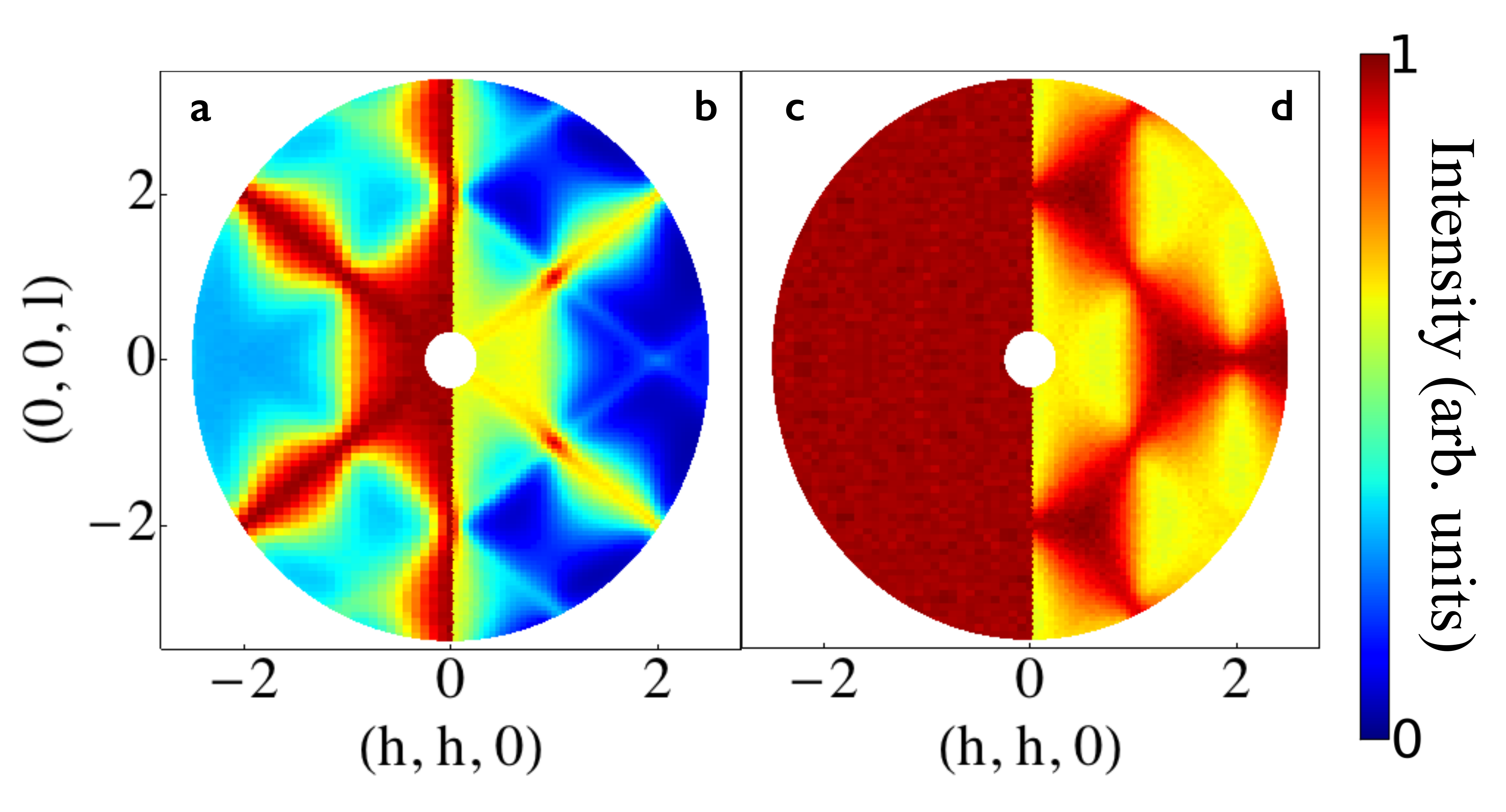}%
\caption{
{\bf Comparison of correlations in spin ice
with those of the spin liquid discussed in this work}
Predictions for polarised neutron scattering experiments 
are shown in the (a-b) spin--flip (SF) and (c-d) non spin--flip (NSF)
channels, as measured by Fennell et al. \cite{fennell09, fennell12}.
(a) Prediction for spin ice in the SF channel exhibiting
 pinch--point singularities.
(b) Prediction for scattering in the SF channel in the spin liquid discussed 
in this work.
(c) Prediction for spin ice in the NSF channel.
This channel is completely featureless in a nearest neighbour
model for spin ice--as shown here--and develops smooth
maxima at the zone boundaries in the presence of long range
dipole interactions \cite{fennell09}.
(d) Prediction for scattering in the NSF channel in the spin liquid discussed 
in this work.
In contrast to spin ice, the spin liquid discussed here 
exhibits singular features in both
SF and NSF scattering.
Results are taken from classical Monte Carlo simulation
of the nearest neighbour model
$\mathcal{H}_{\text{ex}}$ [Eq.~(\ref{eq:Hex})], as described in the text.
}
\label{fig:splitscreen}
\end{figure}

The use of a gauge theory to describe the fluctuations of a spin liquid is
exemplified by the case of 
the spin ice materials R$_2$M$_2$O$_7$ (R=Ho, Dy, M=Ti, Sn)
\cite{castelnovo08, castelnovo12}. 
At low temperatures, the spin configurations in
a spin ice are subject to a constraint
directly analogous to Gauss' law for a magnetic field and consequently may
be described in terms of a gauge theory.
Among the many striking consequences of this is the observation of pinch-point
singularities in the magnetic neutron
scattering structure factor \cite{henley05}, as observed in Ho$_2$Ti$_2$O$_7$~\cite{fennell09}, 
cf. Fig. \ref{fig:splitscreen}(a).
Pinch-point scattering has also been observed in the putative quantum
spin ice Tb$_2$Ti$_2$O$_7$
\cite{fennell12, guitteny13, fritsch13}.
However, in this case
the experimental scattering shows pronounced butterfly-like
features in the non spin-flip (NSF) channel and the 
scattering in the spin-flip (SF) channel
shows narrow arm-like features extending along the $\langle 111 \rangle$
directions of reciprocal space,
neither of which features are predicted for a spin ice.
This raises the question of whether other types of spin liquid may be found
amongst rare-earth pyrochlore magnets.


Here we introduce a different kind of spin liquid on the pyrochlore lattice.
This spin liquid arises on the phase diagram of a realistic model for pyrochlore magnets.
As with spin ice, the theory of this spin liquid contains a gauge
symmetry.
The nature of this theory is fundamentally different to the Maxwellian
theory which describes spin ice, but
just as the emergent gauge structure of spin ice reveals itself in
pinch-point scattering, so the gauge structure of this spin liquid has striking
consequences for scattering experiments.
We will show that at low temperatures, this gauge structure
leads to line-like singularities along the $\langle 111 \rangle$
directions of reciprocal space, which we dub ``pinch lines'' since
they are extended versions of the pinch points exhibited in spin ice.
This is particularly interesting in the light of neutron scattering
results on the pyrochlore magnets Tb$_2$Ti$_2$O$_7$ and 
Yb$_2$Ti$_2$O$_7$ which show strong, sharpening
features along the $\langle 111 \rangle$ directions of
reciprocal space.
Indeed, our theory is able to account for several features
of the diffuse scattering observed in Tb$_2$Ti$_2$O$_7$ \cite{fennell12, fritsch13, guitteny13}, which
are unaccounted for by a theory based on a spin ice model.
\\
\\ 
\\
\\
\\
\noindent {\bf Results} \\

\noindent{\it Spin-liquid regime in a model for pyrochlore magnets}


\begin{figure}[t]
\includegraphics[width=\columnwidth]{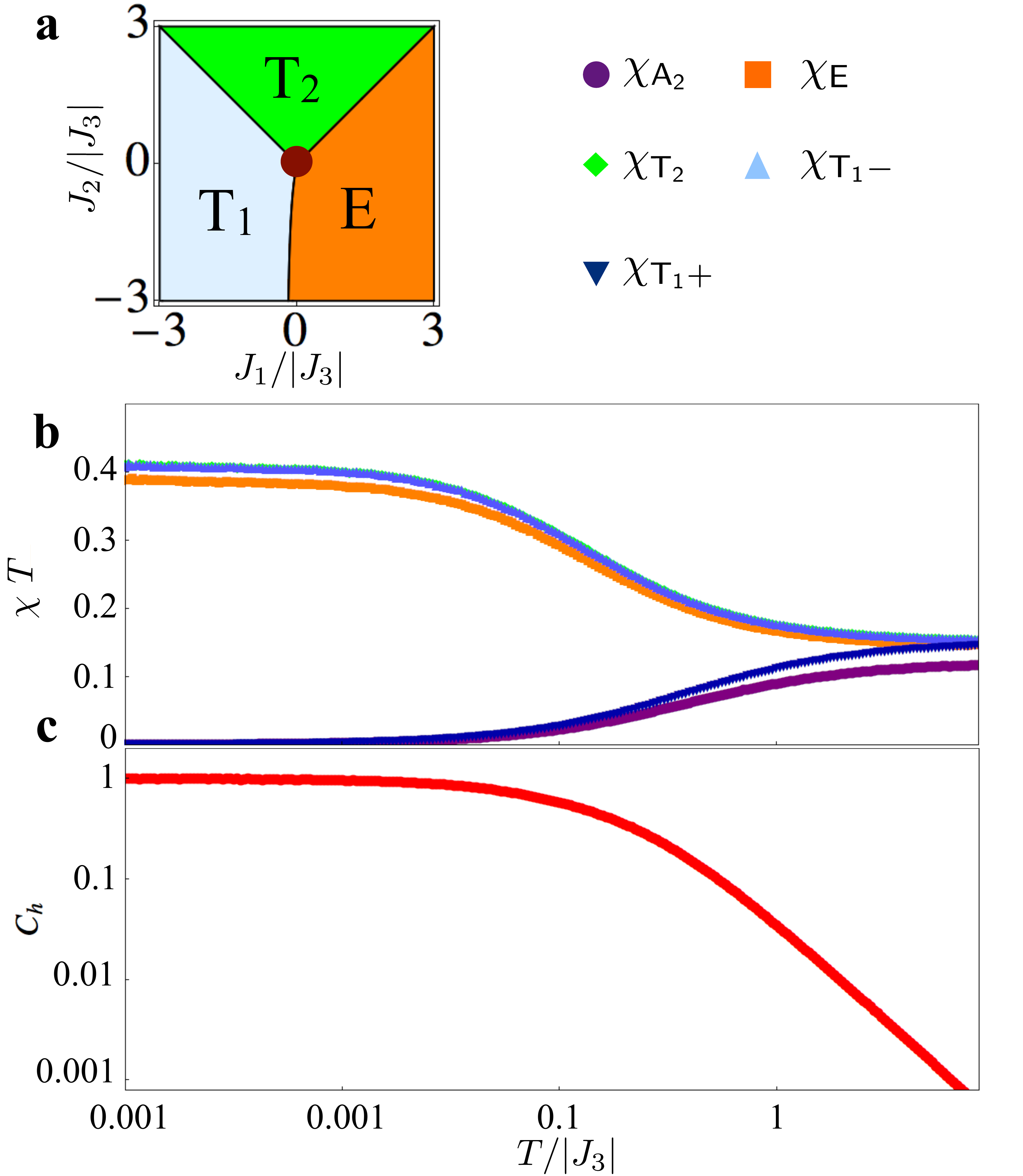}%
\caption{
{\bf Evidence of spin--liquid behaviour from Monte Carlo simulation}.
(a) Classical ground state phase diagram of $\mathcal{H}_{\text{ex}}$
[Eq.~(\ref{eq:Hex})] for $J_3$$<$$0$, in the plane $J_4$$=$$0$, showing 
how ordered phases with symmetry ${\text T_1}, {\text E}$ and ${\text T_2}$ 
meet at the point $J_1$$=$$J_2$$=$$0$ \cite{yan-arXiv}.   
(b) Order--parameter susceptibilities and (c) heat capacity  
calculated in classical Monte Carlo simulation 
of $\mathcal{H}_{\text{ex}}$ [Eq. (\ref{eq:Hex})], 
for parameters $J_1$$=$$J_2$$=$$J_4$$=$$0$, $J_3<0$.  
No phase transition is observed down to $T = 0.001 |J_3|$.  
Instead, the order parameter susceptibilities
of neighbouring ordered phases exhibit a Curie--law crossover, 
characteristic of a Coulombic spin liquid \cite{jaubert13}.
The symbols used for different symmetry channels are shown in an inset.
}
\label{fig:MCthermodynamics}
\end{figure}


We begin with the most general, symmetry-allowed, Hamiltonian for nearest neighbour anisotropic
exchange on the pyrochlore lattice~\cite{curnoe07,ross11-PRX1, yan-arXiv}:
\begin{eqnarray}
\mathcal{H}_{\text{ex}} 
= \sum_{\langle ij \rangle} {\bf S}_i \cdot \bar{\mathcal{J}}_{ij} \cdot {\bf S}_j \ \ , 
\qquad  
\bar{\mathcal{J}}_{01}
= \begin{pmatrix}
J_2 & J_4 & J_4 \\
-J_4 & J_1 & J_3 \\
-J_4 & J_3 & J_1
\end{pmatrix}
\label{eq:Hex}
\end{eqnarray}
where the exchange matrix $ \bar{\mathcal{J}}_{01}$ couples nearest neighbours
along the ${\bf r}_{01}=(0, 1, 1)$ direction and the other exchange matrices 
can be generated from $ \bar{\mathcal{J}}_{01}$ using point group operations.
As shown in \cite{yan-arXiv, benton-thesis}, it is possible to map out the entire classical
ground state phase diagram of Eq. (\ref{eq:Hex}) by an exact transcription of
the Hamiltonian in terms of local fields defined on each
tetrahedron \cite{yan-arXiv, benton-thesis, mcclarty09}:
\begin{eqnarray}
&&\mathcal{H}_{\text{ex}}=
\frac{1}{2}
\sum_{\text{tet}}
\bigg[
\Delta_{\text A_2} m_{\text A_2}^2+
+
\Delta_{\text E} {\bf m}_{\text E}^2
+
\Delta_{\text T_2} {\bf m}_{\text T_2}^2
+ \nonumber \\
&& \qquad \qquad \qquad
\Delta_{\text T_1-} {\bf m}_{\text T_1 -}^2
+
\Delta_{\text T_1+} {\bf m}_{\text T_1 +}^2
\bigg]
+
E_0 
\label{eq:tet-decomp}
\end{eqnarray}
where all the  coefficients $\Delta_{\alpha}\geq0$, $E_0$ is the
ground state energy and the sum runs over all tetrahedra in the
lattice.

The five fields 
$m_{\text A_2}, {\bf m}_{\text E}, {\bf m}_{\text T_2}, {\bf m}_{\text T_1 -}, {\bf m}_{\text T_1 +}$
appearing in Eq. (\ref{eq:tet-decomp}) are defined in Supplementary Table 1.
They transform according to the 
${\text A_2}, {\text E}, {\text T_2}, {\text T_1}$ irreducible representations 
of the point group and have respective dimension $1, 2, 3, 3$ and $3$.
Along a line of points in parameter space the three ordered phases which respectively
maximise the fields ${\bf m}_{\text T_1-}, {\bf m}_{\text T_2}, {\bf m}_{\text E}$ 
become degenerate.
This line includes the point $J_1=J_2=J_4=0, J_3<0$ [cf. Fig. \ref{fig:MCthermodynamics}(a)].
For parameter sets along this line of points
we have $\Delta_{\text A_2}=\Delta_{\text E}=\Delta_{\text T_1-}=0, 
\Delta_{\text A_2} , \Delta_{\text T_1+} > 0$ and
 the Hamiltonian is given by
\begin{eqnarray}
&&\mathcal{H}_{\text{ex}}^{\text{CL}}=
\frac{1}{2}\sum_{\text{tet}}
[\Delta_{\text A_2} m_{\text A_2}^2+
\Delta_{\text T_1+} {\bf m}_{\text T_1 +}^2]+E_0.
\label{eq:HACL}
\end{eqnarray}
%


In a classical ground state of Eq. (\ref{eq:HACL}) it must be the case that
\begin{eqnarray}
m_{\text A_2}=0, \qquad {\bf m}_{\text T_1 +}=0
\label{eq:GSconstraints-lattice}
\end{eqnarray}
for every tetrahedron in the lattice.
All of the results derived in this paper flow
from the implementation of these constraints. 
These provide an exact description of the classical ground states
along the line in parameter space where the three phases
in Fig. \ref{fig:MCthermodynamics}(a) are degenerate.
Observing the consequences of these constraints does
not, however, require precise fine tuning of the
Hamiltonian to Eq. (\ref{eq:HACL}).
These constraints will also dominate the 
physics at finite temperatures for any choice of parameters 
where $\Delta_{\text A_2}, \Delta_{\text E},\Delta_{\text T_1-}<<
\Delta_{\text A_2},
\Delta_{\text T_1+}$
such that energy cost of having a finite value of the
fields ${\bf m}_{\text T_1-}, {\bf m}_{\text T_2}, {\bf m}_{\text E}$ 
is much lower than the cost to have a finite value
of $m_{\text A_2}, {\bf m}_{\text T_1 +}$.

The constraints in Eq. (\ref{eq:GSconstraints-lattice})
are insufficient to select an ordered ground state 
in themselves.
In such circumstances, fluctuations may select a preferred ordered state via the 
order--by--disorder mechanism, but Monte Carlo simulations
indicate that they fail to do so, down to temperatures 3 orders of magnitude below
the scale of the bare coupling [see Fig. \ref{fig:MCthermodynamics}(b)-(c)].
The system thus remains in a disordered but highly-correlated state 
down to low temperature. \\


\noindent{\it Theory of the spin-liquid regime}\\

We can understand the correlations of the spin liquid from Eq. (\ref{eq:GSconstraints-lattice}).
The demand that the fields $m_{\text A_2}$ and ${\bf m}_{\text T_1 +}$
vanish everywhere leaves the fields
$\{
{\bf m}_{\text E}, {\bf m}_{\text T_2}, {\bf m}_{\text T_1 -} \}
$
with freedom to fluctuate in the ground state.
The spatial variation of these fluctuations is constrained by the fact that
neighbouring tetrahedra share a spin, therefore a fluctuation of the local
fields on one tetrahedron affects the values of the 
local fields on the neighbouring tetrahedra.
The fields 
${\bf m}_{\text E}, {\bf m}_{\text T_2}, {\bf m}_{\text T_1 -}$ must therefore
fluctuate in a correlated manner in order to avoid inducing
violations of Eq.~(\ref{eq:GSconstraints-lattice}).
In what follows we show how these correlated fluctuations
can be understood in terms of the fluctuations of a tensor
field with a continuous gauge freedom.


The constraints on the spatial variation of ${\bf m}_{\text E}, {\bf m}_{\text T_2}, {\bf m}_{\text T_1 -}$ 
may be obtained 
from the continuity of fields between $A$ and $B$ sublattice tetrahedra.
The ground state constraints [Eq. (\ref{eq:GSconstraints-lattice})]
in fact imply a set of local conservation laws, on the lattice.
These conservation laws in turn suggest that a coarse-graining approach can
be successful in describing the fluctuations of 
${\bf m}_{\text E}, {\bf m}_{\text T_2}, {\bf m}_{\text T_1 -}$.
And, unlike the global conservation laws which underpin hydrodynamic
theories, these fully local conservation laws
can have consequences even for short
wavelength fluctuations, as we shall see.
Expanding the local constraints to leading order in a gradient
expansion we find
\begin{eqnarray} 
&&\nabla \cdot {\bf m}_{\text T_1 -}=0
\label{eq:A2constraint}
\\
&&
\begin{pmatrix}
\partial_x m_{\text E}^1 \\
-\frac{1}{2} \partial_y m_{\text E}^1 + \frac{\sqrt{3}}{2} \partial_y m_{\text E}^2 \\
-\frac{1}{2} \partial_z m_{\text E}^1 - \frac{\sqrt{3}}{2} \partial_z m_{\text E}^2
\end{pmatrix}
+
\frac{\sqrt{3}}{2} \nabla \times {\bf m}_{\text T_2} \nonumber \\
&& \qquad \quad
-\frac{3}{2} \sin(\phi_{\text T_1}')
\begin{pmatrix}
\partial_y m_{\text T_1-}^z+ \partial_z m_{\text T_1-}^y \\
\partial_z m_{\text T_1-}^x+ \partial_x m_{\text T_1-}^z \\
\partial_x m_{\text T_1-}^y+ \partial_y m_{\text T_1-}^x
\end{pmatrix}=0
\label{eq:T1constraint}
\end{eqnarray}
where the angle $\phi_{\text T_1}'$ is a function of the exchange
parameters, defined in the Methods section.


We wish
to resolve the constraints (\ref{eq:A2constraint})-(\ref{eq:T1constraint}) 
naturally
using a gauge-theoretic approach.
Note that since ${\bf m}_{\text T_1 -}$ appears in both constraints, we cannot
simply introduce separate gauge fields
to resolve Eqs.  (\ref{eq:A2constraint})-(\ref{eq:T1constraint}).
Instead, we incorporate the eight components
of $\{ {\bf m}_{\text E}, {\bf m}_{\text T_2}, {\bf m}_{\text T_1-} \}$ 
into a traceless tensor field $\mathcal{B}$:
\begin{widetext}
\begin{eqnarray}
\mathcal{B}= 
\begin{pmatrix}
m_{\text E}^1 & -\frac{\sqrt{3}}{2} m_{\text T_2}^z - \frac{3 \sin(\phi_{\text T_1}')}{2} m_{\text T_1 -}^z& \frac{\sqrt{3}}{2} m_{\text T_2}^y- \frac{3 \sin(\phi_{\text T_1}')}{2} m_{\text T_1 -}^y \\
\frac{\sqrt{3}}{2} m_{\text T_2}^z - \frac{3 \sin(\phi_{\text T_1}')}{2} m_{\text T_1 -}^z& -\frac{1}{2} m_{\text E}^1 + \frac{\sqrt{3}}{2} m_{\text E}^2& -\frac{\sqrt{3}}{2} m_{\text T_2}^x- \frac{3 \sin(\phi_{\text T_1}')}{2} m_{\text T_1 -}^x\\
-\frac{\sqrt{3}}{2} m_{\text T_2}^y- \frac{3 \sin(\phi_{\text T_1}')}{2} m_{\text T_1 -}^y & \frac{\sqrt{3}}{2} m_{\text T_2}^x
- \frac{3 \sin(\phi_{\text T_1}')}{2} m_{\text T_1 -}^x
&-\frac{1}{2} m_{\text E}^1 - \frac{\sqrt{3}}{2} m_{\text E}^2
\end{pmatrix}
\label{eq:fluxmatrix}
\end{eqnarray}
\end{widetext}


Satisfaction of Eq. (\ref{eq:T1constraint}), along with the condition $\text{Tr}[\mathcal{B}]=0$
is guaranteed by the introduction of a symmetric, tensor field $\mathcal{Y}$
and writing 
\begin{eqnarray}
\mathcal{B}=\mathscr{D} \cdot \mathcal{Y}, \qquad 
\mathscr{D}\equiv
\begin{pmatrix}
0 & -\partial_z & \partial_y \\
\partial_z & 0 & -\partial_x \\
-\partial_y & \partial_x & 0.
\end{pmatrix}
\end{eqnarray} 


The form of the matrix gauge field $\mathcal{Y}$
is then constrained by
Eq. (\ref{eq:A2constraint})
which is satisfied if we take $\mathcal{Y}$ of the form
\begin{eqnarray}
\mathcal{Y}=\begin{pmatrix}
\psi & W_z & W_y \\
W_z & \psi & W_x \\
W_y & W_x & \psi
\end{pmatrix}.
\label{eq:gaugefixedY}
\end{eqnarray}
We can generate alternative forms of $\mathcal{Y}$ by applying Abelian 
gauge transformations
to Eq. (\ref{eq:gaugefixedY}) 
of the form
\begin{eqnarray}
\mathcal{Y}_{\mu \nu}\to\mathcal{Y}_{\mu \nu}+ \partial_{\mu} \partial_{\nu} \zeta 
\label{eq:GT}
\end{eqnarray}
The transformations of Eq. (\ref{eq:GT}) leave the flux matrix ${\mathcal{B}}$, and therefore
the physical spin system, unchanged.
The form of $\mathcal{Y}$ in Eq. (\ref{eq:gaugefixedY}) thus corresponds to a specific choice of gauge.
The theory of the spin liquid is therefore invariant under a group of
gauge transformations $\zeta \in \mathbb{R}$.
Abelian gauge transformations of a similar form to Eq. (\ref{eq:GT}),
acting on tensor fields also appear in the linearized theory of general
relativity \cite{misner73} and the theory of $S=2$ gauge fields \cite{fronsdal78}.


At low temperatures, where there are only fluctuations of the local
fields ${\bf m}_{\text E}, {\bf m}_{\text T_2}, {\bf m}_{\text T_1 -}$, the
free energy will be controlled by the entropy of these fluctuations.
Coarse graining over some volume much larger than a unit cell but
much smaller than the whole system, there will be more states available (and therefore more entropy)
with small values of ${\bf m}_{\text E}, {\bf m}_{\text T_2}, {\bf m}_{\text T_1 -}$ \cite{henley10}.
The most general symmetry-allowed Gaussian free energy describing small fluctuations of these
fields, when written in terms of the tensor field $\mathcal{Y}$, takes the form
\begin{eqnarray}
&&\mathcal{F}_{\text{SL}}=
T \int \frac{d^3 \mathbf{r}}{V_{\text u.c.}}
\bigg[ \lambda_{\text a}
\text{Tr}[(\mathscr{D} \cdot \mathcal{Y})\cdot(\mathscr{D} \cdot \mathcal{Y})^T] \nonumber \\
&& \qquad
+   \lambda_{\text b}
\text{Tr}[(\mathscr{D} \cdot \mathcal{Y})\cdot(\mathscr{D} \cdot \mathcal{Y})] 
+ \lambda_{\text c} \sum_\mu \left(\mathscr{D} \cdot \mathcal{Y} \right)_{\mu \mu}^2 
\bigg] \nonumber \\
\label{eq:effectivetheory1-general}
\end{eqnarray}
which is invariant under the gauge transformations of Eq. (\ref{eq:GT}).\\


\begin{figure*}
\centering
\includegraphics[width=\textwidth]{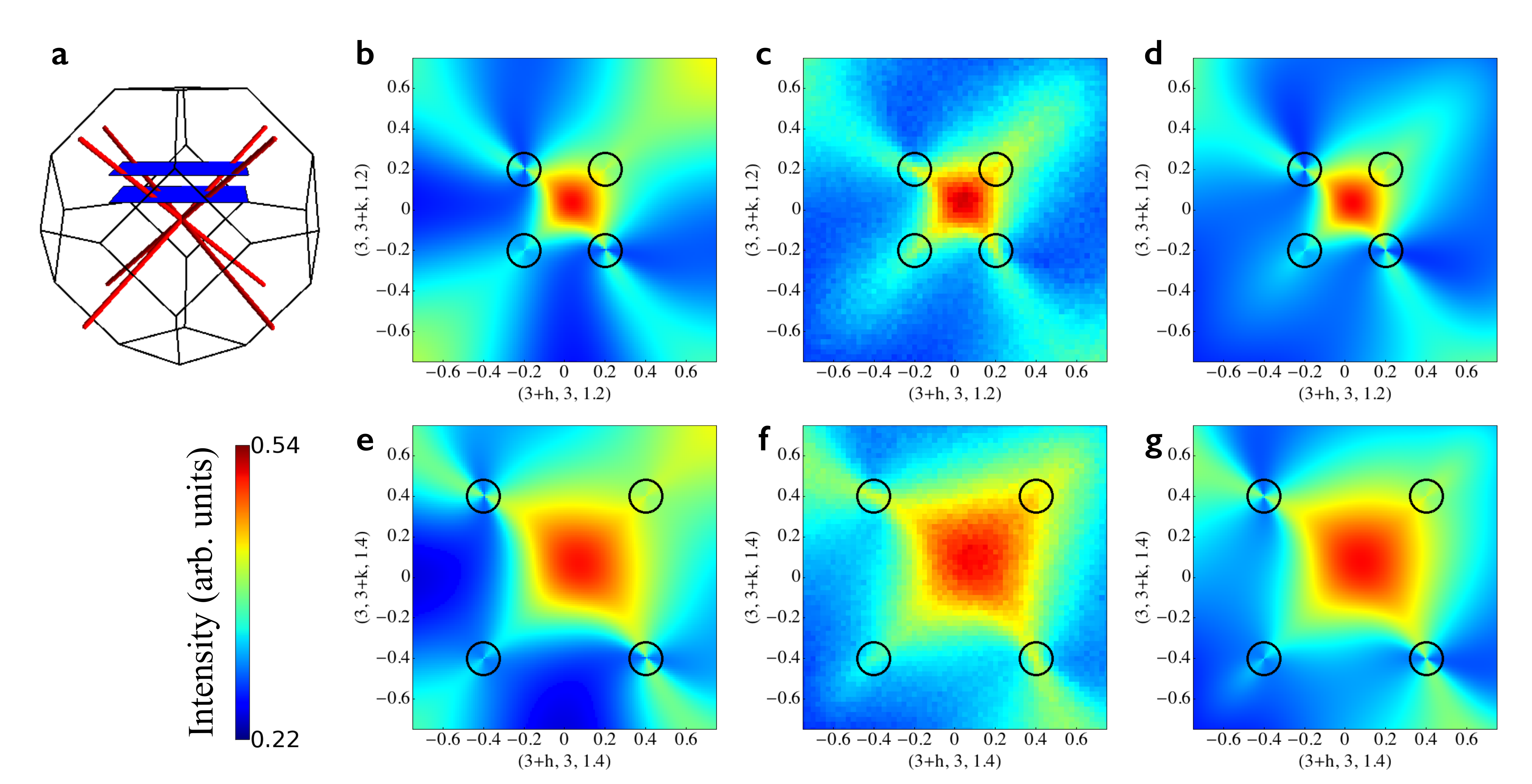}%
\caption{
{\bf Gauge structure of the spin liquid as revealed through pinch--line singularities.}
(a)~Location of pinch--line singularities in reciprocal space.
(b),~(e)~spin structure--factor $S({\bf q})$ in parallel planes in reciprocal space in the
Brillouin zone centred on ${\bf K}=(3, 3, 1)$, as calculated from the continuum
theory [Eq. (\ref{eq:effectivetheory1-general})]. 
Singular features are visible where these planes intersect $\langle111\rangle$ directions,
as indicated by the black circles in each panel.   
These pinch--line singularities, Eq.~(\ref{eq:pinchlineform}), 
are characteristic of the gauge structure of the spin liquid. 
(c),~(f)~spin structure factor calculated in finite temperature Monte Carlo
simulation, in the same regions of reciprocal space.
The pinch lines appear in the simulation results as sharp features around the point where 
$\langle111\rangle$ directions intersect the plane.
(d),~(g)~a calculation of the structure factor made with a lattice based
$1/\mathcal{N}$ theory, described also exhibiting pinch-line singularities.
The simulations were performed at a temperature $T=0.001$ K for
a cluster of $N=256000$ sites and dimensions $40 a_0 \times 40 a_0 \times 10 a_0$
where $a_0$ is the linear size of a cubic unit cell.
Results were calculated 
for parameters $J_1$$=$$0.042$, $J_2$$=$$0.122$, $J_3$$=$$-0.118$, 
$J_4$$=$$-0.04$$\text{\ meV}$, with anisotropic g--tensor $g_\perp/g_{\parallel}$$=$$1/3$, 
in approximate correspondence to Tb$_2$Ti$_2$O$_7$~\cite{cao09}.
Since the crystal field ground state in Tb$_2$Ti$_2$O$_7$
 is a non-Kramers doublet, the finite value of $g_{\perp}$
should be thought of as coming from mixing with the low lying crystal field excitation.
}
\label{fig:pinchline}
\end{figure*}


\noindent{\it Consequences for neutron scattering experiments}\\

The distinctive nature of this spin liquid, and of the theory which describes it [Eq.
(\ref{eq:effectivetheory1-general})], can be revealed
by neutron scattering experiments.
This can be seen by calculating the correlation functions of the local fields 
$\bf m_{\text E}, {\bf m}_{\text T_2}, {\bf m}_{\text T_1 -}$ in momentum space.
In addition to displaying pinch point singularities at zone centres,
these correlation functions are singular approaching any momentum ${\bf q}$
which is along the $(h, h, h)$ directions of reciprocal space, or along any
direction related by the lattice symmetry to $(h, h, h)$.
This contrasts with the case of the Coulombic spin liquid which occurs in the
case of spin ice, where the correlation functions are only singular at the Brillouin zone centre.
Since the fields $\bf m_{\text E}, {\bf m}_{\text T_2}, {\bf m}_{\text T_1 -}$ are simply
linear combinations of the spins, this singular behaviour will also
show up in the spin structure factor $S(\mathbf{q})$, measurable in neutron scattering
experiments.


In the vicinity of one of these singularities, at $T=0$, the scattering is approximated by
\begin{eqnarray}
&&S(\mathbf{K}+\mathbf{q}_{\parallel}+\mathbf{q}_{\perp})
\approx 
\sum_{\alpha \beta}
\gamma_{\alpha \beta} (\mathbf{K}, \mathbf{q}_{\parallel})
\frac{q_{\perp}^{\alpha}q_{\perp}^{\beta}}
{\mathbf{q}_{\perp}^2} 
\label{eq:pinchlineform}
\end{eqnarray}
where $\mathbf{K}$ is a reciprocal lattice vector, $\mathbf{q}_{\parallel}$ is
parallel to a $\langle 111 \rangle$ direction and $\mathbf{q}_{\perp}$
is orthogonal to that direction.
The coefficients $\gamma_{\alpha \beta}$ determine the orientation of the singularity
in $\bf q$-space.
Their dependence on the Brillouin zone ${\mathbf{K}}$ may be thought of as
a form factor determining the contribution of the 
fluctuations of each field ${\bf m}_{\lambda}$ to the scattering in each
Brillouin zone.
The dependence on ${\bf q}_{\parallel}$ is smooth and near a zone center $\mathbf{K}$
one may write $\gamma_{\alpha \beta} (\mathbf{K}, \mathbf{q}_{\parallel})\approx
\gamma_{\alpha \beta} (\mathbf{K}, \mathbf{0})
$.


For $\gamma_{\alpha \beta} \  \cancel{\propto} \ \delta_{\alpha \beta}$ the
structure factor in the limit
 $q_{\perp} \to 0$ will depend on the direction
of approach and we have a singularity, along the entire $\langle 111 \rangle$
direction.
Eq. (\ref{eq:pinchlineform}) has the form of a pinch-point singularity extended
into a line. We therefore will refer to it as a ``pinch-line'' singularity.


These pinch lines can be observed
by taking planar cuts through the scattering, which intersect these lines
away from reciprocal lattice vectors [Fig.~\ref{fig:pinchline}(a)].
This is illustrated using a $T=0$ 
calculation of $S(\mathbf{q})$ from the continuum 
theory [Eq. (\ref{eq:effectivetheory1-general})] ]
in Fig.~\ref{fig:pinchline}(b), (e).
For comparison, we show in Fig.~\ref{fig:pinchline}(c), (f),  the same quantity calculated at
finite temperature within classical Monte Carlo simulation.


The simulation results show sharp features in the structure factor 
approaching the $\langle 111 \rangle$  directions,
as predicted by the theory [Eq. (\ref{eq:GT})].
There is a small broadening of these singularities, coming from the 
finite temperature thermal fluctuations present in the Monte Carlo
simulation.
These features are even more clearly visible in the correlation functions of the 
local fields $\{{\bf m}_{\text E}, {\bf m}_{\text T_2}, {\bf m}_{\text T_1-}  \}$,
-- see Supplementary Figure 1 and Supplementary Note 1.
The presence of the pinch lines in the simulation results is a strong
validation of our theory of the spin liquid regime. 


The continuum theory [Eq. (\ref{eq:GT})] was derived from local constraints,
with associated local conservation laws, and the structure of the theory
is inherited from the structure of those local constraints.
This leads us to expect that the
pinch-line singularities will be robust features of the
spin liquid, even at short wavelengths.
We have confirmed this expectation using two independent,
lattice-based calculations.
Firstly, the sharpening of the scattering around the 
$\langle 111 \rangle$ directions is clearly seen in the
Monte Carlo simulations
in \mbox{Fig.~\ref{fig:pinchline}(c), (f)}.
Secondly, we have also performed
a $1/\mathcal{N}$ calculation of the spin correlations
along the lines of that performed for the Heisenberg model in Ref. \cite{isakov04}.
This calculation 
also predicts pinch-line singularities along the $\langle 111 \rangle$ directions
of reciprocal space, as shown in Fig.~\ref{fig:pinchline}(d), (g).
It is therefore apparent that these singularities are a robust feature of the
spin liquid, arising from the structure of its ground-state constraints, which is
captured by the continuum theory derived in this work.
\\

\noindent {\bf Discussion} \\


Thus far we have uncovered a spin liquid described by a tensor  field
carrying a continuous gauge symmetry,
arising in a particular limit of a realistic model for magnetism on the pyrochlore
lattice [Eq. (\ref{eq:Hex})]. 
The signal feature of this spin liquid is sharp line-like singularities along $\langle 111 \rangle$
directions of reciprocal space, which occur in addition to pinch point singularities at zone centers.
These pinch-line singularities are unique to the spin liquid 
discussed in this paper and as such provide a very discriminating
smoking-gun signature of this novel magnetic state.
In the light of this discovery it is interesting to consider two known pyrochlore materials, which
are often discussed in the context of spin liquid physics: Tb$_2$Ti$_2$O$_7$ and
Yb$_2$Ti$_2$O$_7$.


Tb$_2$Ti$_2$O$_7$ has long been a focal point for discussion of 3-dimensional 
spin liquid physics
\cite{gardner99, molavian07, gardner10}.
While Eq. (\ref{eq:Hex}) alone may not
constitute a complete quantitative model for
the physics of Tb$_2$Ti$_2$O$_7$
it is interesting to compare observations 
on Tb$_2$Ti$_2$O$_7$ with the phenomenology
of the spin liquid.
Polarized neutron scattering experiments on Tb$_2$Ti$_2$O$_7$  have
shown evidence of singular scattering at Brilllouin zone centers,
but the form of this scattering looks rather different to a typical spin ice,
especially in the non-spin flip (NSF) channel.
At the same time, the data presented in Ref. \onlinecite{fennell12}, shows bright, narrow
features extending along the $\langle 111 \rangle$ directions.


As a point of comparison to these experiments, 
the behaviour of the structure factor $S(\mathbf{q})$ in the spin flip (SF) and non spin flip (NSF)
channels, appropriate to a polarised neutron scattering experiment with
initial polarisation ${\bf n} || (1,-1, 0)$, is shown 
in Fig. \ref{fig:splitscreen},
for the same set of exchange parameters as in Fig. \ref{fig:pinchline}.
Narrow prominences 
are visible
in the SF channel along the $\langle111\rangle$ 
directions [Fig. \ref{fig:splitscreen}(b)].
There are also pinch points in both channels at Brillouin zone centers.
The distribution, orientation and polarisation dependence of the
pinch points observed in \cite{fennell12} is the same as that in Fig. \ref{fig:splitscreen} (b), (d).
In particular, we are able to reproduce the shape of the features in the NSF channel, something
which cannot be done with a spin ice based description.
The possibility that the theory described in this work could apply to
Tb$_2$Ti$_2$O$_7$
 is lent weight
by a recent attempt at parameterizing a
pseudo-spin Hamiltonian for  
Tb$_{2+x}$Ti$_{2-x}$O$_{7+y}$ \cite{takatsu-arxiv} which
places it close to the three-way phase boundary at which
this spin liquid emerges in our classical treatment. 


Spin liquid behaviour at finite temperature does not rule
out the possibility of an magnetic order at lower temperature.
Indeed, recent experiments have demonstrated the presence of
competing ordering phenomena in Tb$_2$Ti$_2$O$_7$,
with quadrupolar  \cite{taniguchi13, kadowaki15} and short range ordered
antiferromagnetic states \cite{fritsch13, kermarrec-arXiv, guitteny15} being observed depending
on the sample stoichiometry and experimental cooling protocol.
This is consistent with the nature of the spin liquid considered
in our manuscript, which sits at the confluence of
many competing orders.
In particular, we note that
the ground-state manifold of the spin liquid
contains states consistent with the
${\bf q^{\ast}}=(\pm1/2, \pm1/2, \pm1/2)$ order
observed under field cooled conditions \cite{fritsch13, kermarrec-arXiv}.
These states can only be connected to the other states of 
the spin liquid by rotation of an $\mathcal{O}(L)$ 
number of spins, where $L$ is the linear size of the system.
This may suggest
an explanation for the 
sensitivity to how the system is cooled ---
namely that field cooling may drive the system into a state from which 
it is hard to reach the other parts of the ground state manifold.


The combination of spin liquid physics and prominent features in the scattering along ${\bf q}\parallel(1, 1, 1)$ 
is also strongly reminiscent of the discussion surrounding another pyrochlore: Yb$_2$Ti$_2$O$_7$
\cite{bonville04, ross09, thompson11,chang12, yaouanc13}.
Indeed, it has recently been argued that the unusual physics of this material springs from
competition between the ${\text E}$ and ${\text T_1}$ regions of the phase diagram in 
Fig. \ref{fig:MCthermodynamics}(a) \cite{yan-arXiv, jaubert-arXiv}.
In this context it is not unreasonable to imagine that the physics of the paramagnetic 
phase of Yb$_2$Ti$_2$O$_7$ may be influenced by a nearby spin liquid phase of the 
form described here.
This provides an interesting alternative scenario to quantum spin ice physics in that material.


One concern which arises, in any comparison with experiment, is the extent to which 
the validity of this theory depends on detailed, fine--tuning of parameters.
At first sight, this might seem like a serious obstacle, since it is unlikely that any real material 
would exist exactly at the point where three different ordered phase meet.
However, in practice, a moderate detuning of parameters is only likely to be important in 
determining the nature of the competing (classical) ground state.
As long as experiments are carried out in the disordered phase, at a temperature 
such that violations of the constraint, Eq.~(\ref{eq:GSconstraints-lattice}), are rare, 
the long--wavelength physics will still be described by Eq. (\ref{eq:effectivetheory1-general}), 
and pinch--lines can be observed, albeit with a finite width coming from thermal fluctuations.
The robustness of pinch lines against a finite density of thermally--excited violations 
of Eq.~(\ref{eq:GSconstraints-lattice}) is evidenced by our Monte Carlo simulations
[Fig. \ref{fig:pinchline} (c), (f)], 
which incorporate thermal excitations out of the spin liquid ground state manifold.
Thus, at finite temperature, the signature features of the spin liquid, including 
pinch lines, should remain observable for a finite region of parameter space.


Another important question is the way in which quantum fluctuations 
will affect the properties of the spin liquid at low temperartures.
In the one case which is fully explored, quantum spin ice, quantum tunnelling between 
different spin--ice configurations stabilises a quantum spin--liquid ground state with the 
same U(1) gauge structure as the parent, classical spin liquid 
\cite{hermele04,shannon12,benton12, savary12, hao14,mcclarty15}.  
Meanwhile, stronger off-diagonal exchange interactions between individual spins drive the system 
to order at low temperatures \cite{castroneto06,banerjee08, kato15}.   
However in both cases, classical spin--liquid behaviour is still observed over 
long length--scales, at finite temperature.
Similarly, studies of the quantum $S=1/2$ Heisenberg
model on the pyrochlore lattice find similar spin
correlations \cite{canals98, canals00, huang-arXiv}
to those predicted by the gauge theoretic
description of the classical problem \cite{isakov04}.
In the present case, the nature of the ground state in the presence of quantum 
fluctuations is an open problem, with both ordered and quantum spin--liquid phases 
a realistic possibility.
However, the simplest estimate of the effect of quantum fluctuations, within linear 
spin--wave theory, suggests that the ground state is disordered for a finite 
range of parameters around the classically degenerate point \cite{yan-arXiv}.   
And, given the large entropy associated with the classical spin liquid, 
it seems likely that the system will retain many of its operational features 
--- including the extended pinch--lines --- at finite temperature, regardless
of its quantum ground state.  


The theory presented in this work provides a fundamentally
different paradigm to the emergent electromagnetism known from
spin ice and possesses a gauge freedom bearing an intriguing similarity
to that appearing in the linearized theory of general relativity.
This leads to the possibility of a unified theory of classical spin liquids 
on the pyrochlore lattice, and a classification of the above based on their 
associated gauge freedoms and the consequent singularities in their correlation 
functions.
These issues will be explored further elsewhere.


The discovery of a classical spin liquid is also a promising
starting point to search for new quantum spin liquid ground states.
Previous experience suggests that most quantum spin liquids 
are found in models close to a point of high classical degeneracy, 
and in the case of spin ice the same gauge symmetry underpins both the classical
and quantum spin liquid states. 
We therefore hope that this work can open the way for even richer physics to 
be discovered upon the inclusion of quantum fluctuations.


In conclusion, we have demonstrated the existence of an unusual kind of 
classical spin liquid phase on the pyrochlore lattice, described by the fluctuations 
of a tensor field with a continuous gauge freedom.
The nature of this spin liquid
is revealed by pinch-line 
singularities in correlations which
could be observed in neutron scattering experiments.\\

\noindent{\bf Methods}\\

\noindent{\it  Decomposition of Hamiltonian in terms of local fields}

It was shown in Ref. \cite{yan-arXiv} that the generalized model for 
nearest neighbour exchange on the pyrochlore lattice [Eq. (\ref{eq:Hex})] 
may be exactly rewritten in terms of 
local fields, defined on the pyrchlore tetrahedra
\begin{eqnarray}
&&\mathcal{H}_{\text{ex}}
=
\frac{1}{2}
\sum_{\text{tet}}
\big(
\Delta_{\text A_2} m_{\text A_2}^2+
\Delta_{\text E} {\bf m}_{\text E}^2+
\Delta_{\text T_2} {\bf m}_{\text T_2}^2+ \nonumber \\
&& \quad 
\Delta_{\text T_1 +} {\bf m}_{\text T_1 +}^2+
\Delta_{\text T_1 -} {\bf m}_{\text T_1 -}^2
\big) + \text{constant}.
\label{eq:H1}
\end{eqnarray}
The fields are labelled by the irreducible
representations of the point group according to which
they transform.
These fields are defined in Supplementary Table I.

The angle $\phi_{\text T_1}'$ which appears in the definitions of 
${\bf m}_{\text T_1 +}$ and ${\bf m}_{\text T_1 -}$ and Eq. (\ref{eq:T1constraint})
is chosen
such that there is no bilinear coupling between 
${\bf m}_{\text T_1 +}$ and ${\bf m}_{\text T_1 -}$
and such that
\begin{eqnarray}
a_{\text T_1 -}\leq a_{\text T_1 +}.
\end{eqnarray}

Note that this convention for the
definition of the $\text T_1$ symmetric fields 
is different to that chosen
in Ref. \cite{yan-arXiv}.\\

\noindent{\it Monte Carlo Simulation}

The classical Monte Carlo simulations used to obtain the results in
Figs. \ref{fig:splitscreen}-\ref{fig:pinchline} 
are based on the Metropolis algorithm 
with parallel tempering~\cite{swendsen86,geyer91} and  over-relaxation~\cite{creutz87}. 
The spins are treated as classical vectors of fixed length $|S_{i}|=1/2$ with local updates
using the Marsaglia method~\cite{marsaglia72}.   

Following common practice in Monte Carlo simulations,
the order parameter susceptibilities appearing in Fig. \ref{fig:MCthermodynamics}
are calculated according to the following formula:
\begin{eqnarray}
\chi=\frac{N}{T} \left( \langle {\bf m}_{\lambda}^2 \rangle -\langle |{\bf m}_{\lambda}| \rangle^2 \right)
\end{eqnarray}
where $N$ is the number of spins in the system, $T$ is the temperature
and ${\bf m}_{\lambda}$ are the local fields
appearing in Eq. (\ref{eq:tet-decomp}).\\

\noindent{\it Lattice based calculation of the structure factor}

For the purposes of comparison with the continuum theory developed in the main text, 
we have also 
performed some lattice based calculations of the correlations in the spin liquid
regime.

These calculations follow a method which has been previously been shown successful
in understanding the correlations of disordered phases of spin ice \cite{henley05} 
the Heisenberg model
on the pyrochlore lattice \cite{isakov04, conlon10}, and 
protons in water ice \cite{isakov15}.

In this approach the constraints on the lengths of the spins 
\begin{eqnarray}
{\bf S}_i^2=S^2
\label{eq:hardconstraint}
\end{eqnarray}
are only enforced on average  
\begin{eqnarray}
\langle{\bf S}_i^2\rangle=S^2.
\label{eq:aveconstraint}
\end{eqnarray}

Eq. (\ref{eq:aveconstraint}) is enforced by means of a
Lagrange multiplier $\lambda$ added to the Hamiltonian.
We write
\begin{eqnarray}
\beta H \to \beta H_{\lambda}=\beta H + \lambda \sum_i {\bf S}_i^2
\end{eqnarray}
where $\beta$ is the inverse temperature.

Using a Fourier transformation $\beta H_{\lambda}$ may be written as
\begin{eqnarray}
\beta H_{\lambda}=\frac{1}{2} \tilde{S}(-\mathbf{q}) \cdot \mathcal{M}(\mathbf{q}) \cdot 
\tilde{S}(\mathbf{q})
\end{eqnarray}
where is $\tilde{S}(\mathbf{q})$ is a 12-component vector formed from the Fourier transforms
of the 3 spin components on each of the 4 sublattices.

The correlations of $\tilde{S}(\mathbf{q})$ are then
\begin{eqnarray}
\langle \tilde{S}_i (-\mathbf{q}) \tilde{S}_j(\mathbf{q}) \rangle
=
(\mathcal{M}^{-1} (\mathbf{q}))_{ij}
\end{eqnarray}
and $\lambda$ can be chosen such that Eq. (\ref{eq:aveconstraint}) is 
obeyed.

Where $\mathcal{M}(\mathbf{q})$ possesses flat bands of 
eigenvalues at the bottom of its spectrum-- as is the case
in the spin liquid regime-- the limit $T\to0$ of the correlation function
becomes a projection matrix, projecting into the subspace 
described by the associated eigenvectors \cite{isakov15}.
This projection operator can be thought of as enforcing the local
ground state constraints \cite{henley05}.

It is this, zero-temperature, limit of the correlation function which is 
plotted in Figs. 3(d) and 3(g) of the main text.

The approach outlined here can be constructed as a perturbative expansion
in powers of $1/\mathcal{N}$, where $\mathcal{N}$ is a number of
copies of the system and the spin length constraint [Eq. (\ref{eq:aveconstraint})]
becomes
\begin{eqnarray}
\frac{1}{\mathcal{N}}\sum_{\alpha=1}^{\mathcal{N}}{\bf S}_{i, \alpha}^2=S^2
\end{eqnarray}

This method is described in more detail in Ref. \cite{benton-thesis}. \\


\noindent {\bf Acknowledgements} \\

This work was supported by the Theory of Quantum Matter Unit of the 
Okinawa Institute of Science and Technology Graduate University.\newline


\noindent {\bf Author contributions} \\

The analytic calculations were performed by O.~B. and H.~Y.
The Monte Carlo simulations were written and performed by L.~D.~C.~J.
The manuscript was written by O.~B. with input from all authors.
N.~S. supervised the project.\\

\noindent {\bf Competing financial interests} \\

The authors declare no competing financial interests.\\

\noindent {\bf Data availability statement} \\

This is a theoretical work. 
The authors declare that the data supporting the findings of this study
are available within the article and it's supplementary information.
\\



\end{document}